\documentclass[aps,prl,twocolumn,groupedaddress]{revtex4-1}
\usepackage{graphicx,color} 
\usepackage{dcolumn}
\usepackage{bm}
\usepackage{amssymb,amsmath}

\newcommand{\be}{\begin{equation}}
\newcommand{\ee}{\end{equation}}

\begin{document}

\preprint{}
\title{Direct Measurements of Magnetic Polarons in Cd$_{1-x}$Mn$_x$Se Nanocrystals from Resonant Photoluminescence}

\author{W.~D.~Rice$^{1,6}$, W.~Liu$^{2}$, V.~Pinchetti$^{3}$, D. R. Yakovlev$^{4,5}$, V.~I.~Klimov$^{2}$, S.~A.~Crooker$^{1}$}
\affiliation{$^1$National High Magnetic Field Laboratory, Los Alamos National Laboratory, Los Alamos, NM 87545, USA}
\affiliation{$^2$Chemistry Division, Los Alamos National Laboratory, Los Alamos, NM 87545, USA}
\affiliation{$^3$Dipartimento di Scienza dei Materiali, Universit\`a degli Studi di Milano-Bicocca, Via Cozzi 55, IT-20125 Milano, Italy}
\affiliation{$^4$Experimentelle Physik 2, Technische Universit\"at Dortmund, D-44221 Dortmund, Germany}
\affiliation{$^5$Ioffe Institute, Russian Academy of Sciences, 194021 St. Petersburg, Russia}
\affiliation{$^6$Department of Physics and Astronomy, University of Wyoming, Laramie, WY 82071, USA}



\begin{abstract}
In semiconductors, quantum confinement can greatly enhance the interaction between band carriers (electrons and holes) and dopant atoms. One manifestation of this enhancement is the increased stability of \textit{exciton magnetic polarons} in magnetically-doped nanostructures. In the limit of very strong 0D confinement that is realized in colloidal semiconductor nanocrystals, a single exciton can exert an effective exchange field $B_{\rm{ex}}$ on the embedded magnetic dopants that exceeds several tesla. Here we use the very sensitive method of resonant photoluminescence (PL) to directly measure the presence and properties of exciton magnetic polarons in colloidal Cd$_{1-x}$Mn$_x$Se nanocrystals. Despite small Mn$^{2+}$ concentrations ($x$=0.4-1.6\%), large polaron binding energies up to $\sim$26~meV are observed at low temperatures via the substantial Stokes shift between the pump laser and the resonant PL maximum, indicating nearly complete alignment of all Mn$^{2+}$ spins by $B_{\rm{ex}}$. Temperature and magnetic field-dependent studies reveal that $B_{\rm{ex}} \approx$ 10~T in these nanocrystals, in good agreement with theoretical estimates. Further, the emission linewidths provide direct insight into the statistical fluctuations of the Mn$^{2+}$ spins.  These resonant PL studies provide detailed insight into collective magnetic phenomena, especially in lightly-doped nanocrystals where conventional techniques such as nonresonant PL or time-resolved PL provide ambiguous results. 
\end{abstract}
\maketitle


Advances in the colloidal synthesis of magnetically-doped nanomaterials have sparked a renewed focus on low-dimensional magnetic semiconductors \cite{NorrisNL, Mikulec, Hoffman, Erwin, Ithurria, Bhargava, BeaulacAM, Bussian}. The interesting magnetic properties of these materials originates in the strong \textit{sp-d} exchange interactions that exist between carrier spins (\textit{i.e.}, band electrons and holes with \textit{s-} and \textit{p-}type wavefunctions) and the local $3d$ spins of embedded paramagnetic dopant atoms such as Mn, Co, or Fe \cite{FurdynaJAP, Dietl, DMSbook}. At the microscopic level, the strength of this interaction for a dopant located at position $\textbf{r}_i$ scales with the probability density of the carrier envelope wavefunctions at that point: $| \psi_{e,h}(\textbf{r}_i)|^2$. As such, local spin-spin interactions can be greatly enhanced by strong quantum confinement, which compresses carrier wavefunctions to nanometer-scale volumes and therefore increases $| \psi_{e,h}(\textbf{r})|^2$. The extent to which these exchange interactions can be enhanced and controlled via quantum confinement is an area of significant current interest and has recently been studied in a variety of magnetically-doped semiconductor nanostructures, including nanoribbons \cite{Yu, Fainblat}, nanoplatelets \cite{Murphy}, epitaxial quantum dots \cite{Maksimov, Seufert, Bacher, Dorozhkin, Mackowski, Besombes, Wojnar, Abolfath, Sellers, Trojnar, Klopotowski, Kobak}, and colloidal nanocrystals \cite{NorrisNL, Mikulec, Hoffman, Erwin, Ithurria, Bhargava, BeaulacAM, Bussian, Bhattacharjee, BeaulacScience, Nelson, RiceNNANO}.

A particularly striking consequence of \textit{sp-d} interactions in II-VI semiconductors is the formation of \textit{exciton magnetic polarons} (EMPs), wherein the effective magnetic exchange field from a single photogenerated exciton -- $B_{\rm{ex}}$ -- induces the collective and spontaneous ferromagnetic alignment of the magnetic dopants within its wavefunction envelope, generating a net local magnetization even in the absence of any applied field \cite{YakovlevChapter, DietlPRL1995, Kavokin}.  In turn, these aligned local moments act back on the exciton's spin, which lowers the exciton's energy, further localizes the exciton, and further stabilizes the polaron. The stability and binding energy of EMPs therefore depends on the detailed interplay between many factors including the exciton lifetime, the polaron formation time, the exchange field $B_{\rm{ex}}$, sample dimensionality, and temperature.

EMPs and collective magnetic phenomena have been experimentally studied in a variety of Mn$^{2+}$-doped semiconductor nanostructures, including CdMnSe and CdMnTe-based epilayers and quantum wells \cite{Zayhowski, Itoh, Yakovlev1992, Mackh1, Mackh2, Takeyama, Oka, Fiederling, Kusrayev}, self-assembled CdMnSe and CdMnTe quantum dots grown by molecular-beam epitaxy \cite{Maksimov, Seufert, Bacher, Dorozhkin, Mackowski, Besombes, Wojnar, Abolfath, Sellers, Trojnar, Klopotowski}, and most recently in CdMnSe nanocrystals synthesized via colloidal techniques \cite{BeaulacScience, Nelson}. Common measurement techniques include the analysis of conventional (\textit{i.e.}, non-resonant) PL \cite{Maksimov, Bacher, Dorozhkin, Wojnar, Takeyama, BeaulacScience, Nelson} and time-resolved PL \cite{Seufert, Oka, Itoh, Zayhowski, BeaulacScience, Nelson}.

However, an especially powerful and incisive experimental technique for directly revealing the presence and properties of magnetic polarons is the method of \textit{resonant} PL \cite{YakovlevChapter, Mackh1, Mackh2, Itoh, Yakovlev1992, Fiederling, Kusrayev}. Here, a narrow-band excitation laser, tuned to the low-energy side of the exciton absorption peak, resonantly excites low-energy `cold' excitons. Subsequently, the exciton's exchange field $B_{\rm{ex}}$ aligns the Mn$^{2+}$ spins and forms an EMP, which in turn lowers the exciton's energy.  When the exciton recombines, it therefore emits a lower-energy photon. Thus, the Stokes shift between the pump laser energy and the emitted PL directly reveals the polaron binding energy. Temperature and magnetic field dependent measurements determine $B_{\rm{ex}}$, and therefore the strength of \textit{sp-d} interactions in the material. Moreover, the resonant PL linewidth provides immediate insight into Mn$^{2+}$ spin fluctuations.  Resonant PL methods have been used to quantify EMP energies and spin fluctuations in Mn$^{2+}$-doped semiconductor epilayers \cite{Mackh1, Kusrayev} and quantum wells \cite{Yakovlev1992, Mackh2, Fiederling}. To date, however, this powerful technique has never been applied to study EMPs in colloidal nanocrystals, despite the fact that they represent the strongest case of 0D quantum confinement, in which \textit{sp-d} exchange is expected to be most enhanced.

To address this gap, here we use resonant PL spectroscopy to directly reveal the presence and properties of EMPs in Cd$_{1-x}$Mn$_x$Se colloidal nanocrystals. Despite small Mn$^{2+}$ concentrations ($x$=0.4-1.6\%), large EMP binding energies up to $\sim$26~meV are observed at low temperatures, indicating that within its radiative lifetime a photogenerated exciton completely aligns the Mn$^{2+}$ spins within its host nanocrystal. The exchange field $B_{\rm{ex}}$ is found to lie in the range of 8-11~T, in good agreement with theoretical expectations for these nanocrystals. Finally, the significant variations of the resonant PL linewidth are confirmed by numerical Monte Carlo simulations, providing direct insight into Mn$^{2+}$ spin fluctuations. These studies highlight the utility of resonant PL as an effective tool for detailed studies of collective magnetic phenomena in colloidal nanomaterials.

\begin{figure} [tbp]
\includegraphics [width=0.9 \columnwidth] {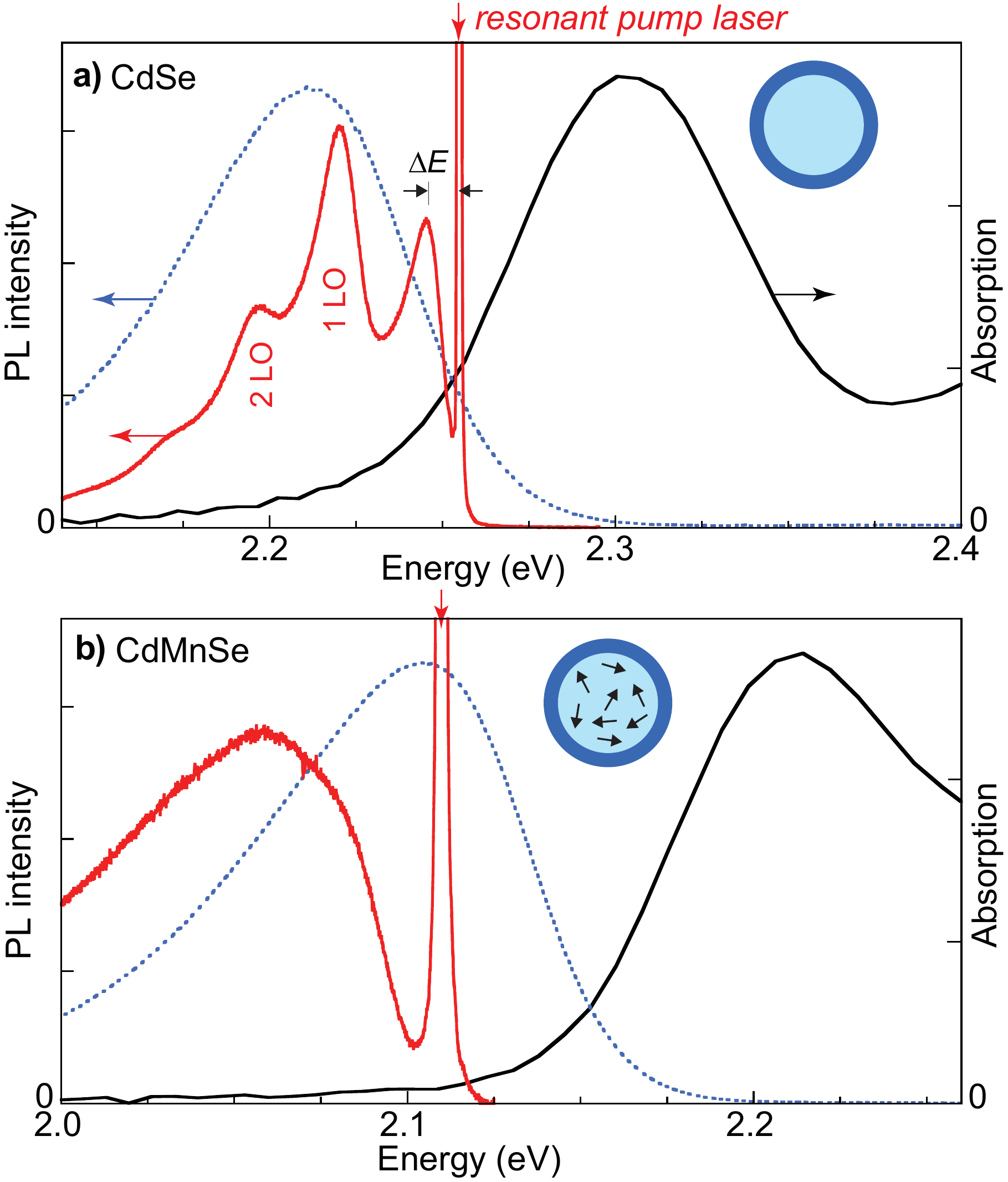}
\caption{(a) Absorption (black), non-resonant PL (dotted blue), and resonant PL (red) from a reference sample of nonmagnetic CdSe nanocrystals at $T=1.8$~K and $B=0$~T.  Under resonant (selective) excitation, the  PL emission exhibits a series of sharp peaks that reveal exciton fine structure and LO-phonon replicas, in agreement with past studies. (b) Absorption, non-resonant PL, and resonant PL from the 1.6\% CdMnSe nanocrystals at 1.8~K and 0~T. In contrast to the nonmagnetic sample, resonant PL from these nanocrystals exhibits a large Stokes shift and is very broad, which provides a direct measure of the magnetic polaron binding energy and also the Mn$^{2+}$ spin fluctuations, respectively.}
\label{Fig1}
\end{figure}

Several batches of Cd$_{1-x}$Mn$_x$Se nanocrystals were grown by colloidal synthesis (see Supporting Information). The nanocrystals were grown sufficiently large (5~nm average diameter) so that the 1S exciton energy lies below the $^4T_1 \rightarrow$$^6A_1$ transition energy of the Mn$^{2+}$ ions. This results in long exciton recombination lifetimes exceeding 10~ns, which is much longer than typical sub-nanosecond polaron formation times \cite{Nelson, Seufert, YakovlevChapter, Zayhowski, Itoh, Oka}.  Polaron formation is therefore not interrupted by exciton recombination, and proceeds to an equilibrium condition.  The Mn$^{2+}$ concentrations (0.4\%, 0.7\%, and 1.6\%, which correspond to $\langle N \rangle \approx$ 5, 9, and 20 Mn$^{2+}$/nanocrystal) were intentionally kept low to minimize Mn$^{2+}$ clustering effects.  Reference samples of nonmagnetic CdSe nanocrystals were also synthesized. The nanocrystals were dispersed in optical-quality polyvinylpyrrolidone films to minimize scattering.

The nanocrystals were first characterized in applied magnetic fields $B$ up to 7~T and at temperatures down to 1.8~K using standard techniques for circularly-polarized absorption, non-resonant PL, magnetic circular dichroism, and time-resolved PL. While certain aspects of these conventional measurements are consistent with EMP formation (see Supporting Information), alternative scenarios are also difficult to rule out: for example, the dynamic spectral redshifts that appear in time-resolved PL studies, which may indicate EMP formation, have also been observed in \textit{non}-magnetic nanocrystals due to energy transfer \cite{CrookerPRL2002, Liu}.

Most importantly, therefore, we focused on \emph{resonant} PL techniques to explicitly detect the presence and properties of EMPs. Here, we tuned a continuous-wave dye laser to the low-energy side of the inhomogeneously-broadened 1S exciton absorption peak.  This selectively pumps excitons into the subset of nanocrystals whose optically-allowed (``bright" exciton) absorption energy is exactly resonant with the laser, thereby mitigating the effects of ensemble broadening. Subsequently, these excitons can lower their energy by relaxing to ``dark" states within the exciton fine structure \cite{NorrisPRB1996, Nirmal} or also -- in Mn$^{2+}$-doped nanocrystals -- by forming an EMP.  The key point is that the resonant Stokes shift $\Delta E$ between the pump laser and the emitted PL therefore reveals the difference between the exciton's initial energy (\textit{i.e.}, at absorption) and final energy (\textit{i.e.}, at recombination).  Resonant PL is closely related to fluorescence line narrowing methods that have been used to observe the bright-dark splitting (fine structure) of excitons in conventional undoped nanocrystals \cite{NorrisPRB1996, Nirmal}.

To establish a reference against which to compare the Mn$^{2+}$-doped nanocrystals, we first show in Fig. 1a the characteristic absorption, non-resonant PL, and resonant PL spectra from an ensemble of conventional nonmagnetic CdSe nanocrystals at low temperature (1.8~K) and at $B$=0. The 1S exciton absorption peak (black trace) is well-defined and $\sim$70~meV wide, indicating a typical degree of inhomogeneous broadening due to the nanocrystals' size distribution. The nonresonant PL (dotted blue trace) is similarly broad and is shifted to lower energies, as expected. The resonant PL spectrum (red trace) exhibits a series of sharper peaks that are Stokes-shifted from the pump laser, and is also quite typical \cite{NorrisPRB1996, Nirmal, Furis}.  Here, the energy difference $\Delta E$ between the laser and the closest emission peak ($\sim$6~meV in this sample) directly reveals the exciton's fine structure -- \textit{i.e.}, the energy splitting $\Delta_{\textrm{bd}}$ between the ``bright" (angular momentum projection $J=\pm 1$) and lower-energy ``dark" ($J=\pm 2$) exciton states that is due to electron-hole exchange. The additional peaks at lower energies are optical phonon replicas of the dark state.

In comparison, Fig. 1b shows the same three spectra measured on the 1.6\% CdMnSe nanocrystals. The broad absorption and non-resonant PL spectra are dominated by the inhomogeneous size distribution and are essentially indistinguishable from those of the nonmagnetic reference sample (the overall shift in absorption/PL energy is because the nonmagnetic nanocrystals in Fig. 1a have smaller average diameter).  Crucially, however, the \emph{resonant} PL spectrum is completely different. Instead of a series of sharp emission peaks, the resonant PL at $B=0$ exhibits a much larger Stokes shift $\Delta E$, and is broad and featureless. This indicates i) that excitons in these magnetically-doped nanocrystals lower their energy much more during their lifetime as compared to nonmagnetic nanocrystals, and ii) that the distribution of Stokes shifts is much broader.  Both of these properties are consistent with the presence of Mn$^{2+}$ spins and the formation of EMPs.

\begin{figure*} [tbp]
\includegraphics [width=4 in] {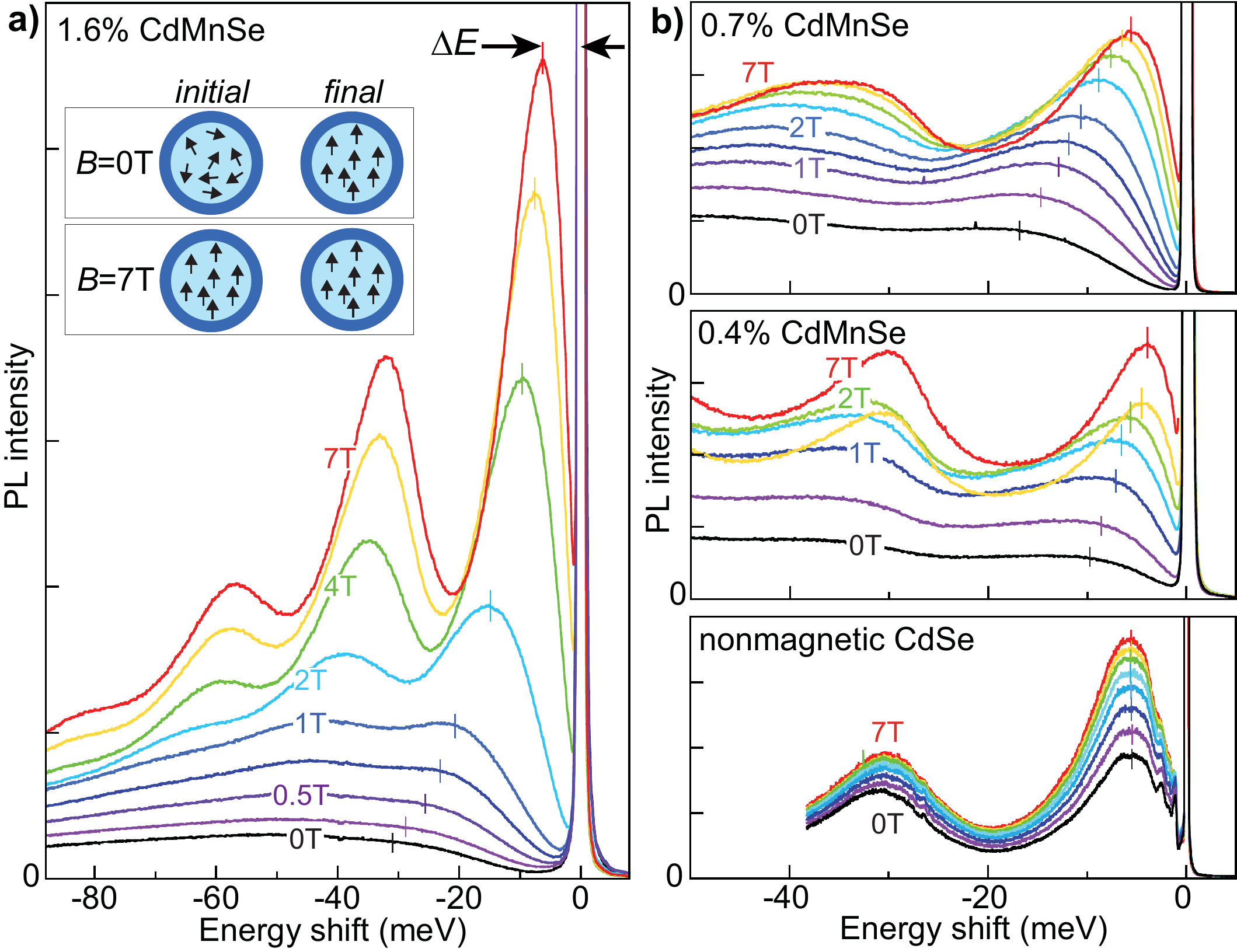}
\caption{(a) Magnetic field dependent resonant PL spectra from the 1.6\% CdMnSe nanocrystals at 1.8~K. With increasing field $B$ from $0 \rightarrow 7$~T (applied in the Faraday geometry), the resonant emission sharpens dramatically and evolves into a narrow peak (and its optical phonon replicas), while the Stokes shift $\Delta E$ between the excitation laser and the first emission peak is markedly reduced. Respectively, these indicate a suppression of Mn$^{2+}$ spin fluctuations, and a reduction of the magnetic polaron binding energy (because the Mn$^{2+}$ are already initially aligned by $B$, and further alignment by the exciton is not possible). The cartoons depict the initial and final Mn$^{2+}$ spin configurations (\textit{i.e.}, when the exciton is created, and when it recombines), at both zero and large $B$. (b) Resonant PL spectra versus $B$ at 1.8~K for the 0.7\% CdMnSe, 0.4\% CdMnSe, and nonmagnetic CdSe nanocrystals. As the Mn$^{2+}$ concentration decreases, the zero-field Stokes shift and the linewidth both decrease (because both the polaron binding energy and Mn$^{2+}$ spin fluctuations decrease). However, by 7~T the resonant PL spectra from all the CdMnSe nanocrystals closely resembles the resonant PL from the nonmagnetic CdSe nanocrystals (narrow peaks, small $\Delta E$). $B$ has little influence on $\Delta E$ or the resonant PL linewidth in nonmagnetic nanocrystals.}
\label{Fig2}
\end{figure*}

The most unambiguous evidence for EMP formation in these CdMnSe nanocrystals is obtained via the dramatic evolution of the resonant PL spectra in applied magnetic fields. As shown in Fig. 2a for the 1.6\% CdMnSe sample, the initially broad and featureless resonant PL spectrum at 0~T evolves continuously into a series of narrow emission peaks at 7~T. The Stokes shift $\Delta E$ between the laser and the closest emission peak decreases from $\sim$30~meV to $\sim$5~meV. Importantly, at 7~T the resonant PL spectrum now closely resembles that which was obtained from the \textit{non}magnetic CdSe nanocrystals (see Fig. 1a); namely, a narrow emission peak at small $\Delta E$ followed by a series of optical phonon replicas. Measurements of the 0.7\% and 0.4\% CdMnSe nanocrystals show similar evolution with field, but with smaller $\Delta E$ at $B=0$ (Fig. 2b). In marked contrast, all the nonmagnetic CdSe nanocrystals show very little change in the resonant PL energy and width up to 7~T, in agreement with prior studies \cite{Nirmal, Furis}.

These two trends in the resonant PL -- namely, the drop in both $\Delta E$ and the linewidth as $B$ is increased -- are telltale hallmarks of EMPs that have been previously observed in higher-dimensional and rather heavily Mn-doped diluted magnetic semiconductors such as CdMnTe-based epilayers and quantum wells with $x >$10\% \cite{Yakovlev1992, Mackh1, Mackh2, Fiederling}. These trends, now clearly observed here in lightly doped but strongly confined 0D nanocrystals, are due to the decrease of the EMP binding energy, and also the decrease of Mn$^{2+}$ spin fluctuations, by the applied field $B$.  As the cartoons in Fig. 2 depict, at $B=0$ the excitons are initially photoinjected into nanocrystals having a randomly-oriented distribution of Mn$^{2+}$ spins (the \textit{average} Mn$^{2+}$ spin polarization in the ensemble is zero). Subsequently, the excitons lower their energy by forming a magnetic polaron, and eventually recombine from a lower-energy final state in which the Mn$^{2+}$ spins are aligned (note this occurs independent of the nanocrystals' orientation).  In these ensemble measurements, this results in a large average Stokes shift $\Delta E$ at $B=0$. Moreover, because of the initially random orientation of the Mn$^{2+}$ spins, and also because the exact number and location of the Mn$^{2+}$ spins within each nanocrystal varies, the energy lost by each exciton is different. This gives a broad distribution of measured Stokes shifts, and correspondingly broad resonant PL features at $B$=0~T.

In contrast, at $B$=7~T (and at low temperatures) each exciton is initially photoinjected into a nanocrystal wherein all the paramagnetic Mn$^{2+}$ spins are \textit{already} aligned by $B$. The exciton cannot further align the Mn$^{2+}$, and so the final energy of the exciton is approximately the same as its initial energy. The net result is a small resonant Stokes shift $\Delta E$ that is expected to be similar to that from nonmagnetic nanocrystals, as observed.  Moreover, because the Mn$^{2+}$ within each nanocrystal are aligned in both the initial and final states (again, independent of the nanocrystals' orientation), there is no additional broadening from stochastic spin fluctuations.  The resonant PL features are expected to be comparably narrow to those obtained from nonmagnetic nanocrystals, again precisely as observed (\textit{e.g.}, compare Fig. 2a and Fig. 1a).

To better quantify the average Stokes shift $\Delta E$ and also the linewidth $\Gamma$ of the resonant PL features, we fit all the resonant PL spectra from all the nanocrystals to a series of Gaussian peaks. The results are shown in Figs. 3a,b. We note that these values do not depend on the exact photon energy of the resonant pump laser, as long as it is tuned well below the 1S exciton (see Supporting Information). At $B$=0, $\Delta E$ is largest for the 1.6\% CdMnSe nanocrystals ($\sim$31 meV), and is smaller for the more lightly-doped samples ($\sim$17 meV for the 0.7\% nanocrystals and $\sim$10 meV for the 0.4\% nanocrystals). Crucially however, for all the Mn$^{2+}$-doped nanocrystals, $\Delta E$ falls monotonically as $B \rightarrow 7$~T, and converges  towards a common value of $\sim$5~meV at large $B$. 

Interestingly, 5~meV is very close to the field-independent value of $\Delta E$ that is observed in nonmagnetic nanocrystals of the same size (black points), which in turn is due to electron-hole exchange and the splitting $\Delta_{\textrm{bd}}$ between bright and dark excitons \cite{NorrisPRB1996}, as discussed above. However, the presence of a similar 5~meV offset exhibited by $\Delta E$ in the CdMnSe nanocrystals is somewhat unexpected, because the exciton ground state in conventional Mn$^{2+}$-doped II-VI semiconductors is typically bright at large $B$ \cite{FurdynaJAP, DMSbook} (this is because the \emph{s-d} electron-Mn$^{2+}$ exchange energy -- the energy required to flip the electron spin -- typically greatly exceeds $\Delta_{\textrm{bd}}$). Our resonant PL data are nonetheless consistent with a lower-lying and/or dark exciton ground state in these CdMnSe nanocrystals at large $B$, whose origin is unclear but which could result, \emph{e.g.}, from slow EMP reorientation \cite{Nelson} or from a confinement-induced reduction/inversion of the \emph{s-d} exchange interaction (a scenario recently explored in quantum-confined semiconductors \cite{Bussian, Yu, Fainblat, Merkulov1999}). Regardless, the essential observation and key point is that $\Delta E$ clearly converges to a constant value of $\sim$5~meV at large $B$. We therefore associate the EMP binding energy (\textit{i.e.}, the energy that is lost by the exciton at $B$=0 due explicitly to EMP formation) with $\Delta E - 5$~meV $\equiv$ 26~meV, 12~meV, and 5~meV for the 1.6\%, 0.7\%, and 0.4\% CdMnSe nanocrystals, respectively.

Similar to the Stokes shifts, the resonant PL linewidths $\Gamma$ are, at $B$=0, largest for the 1.6\% CdMnSe nanocrystals and smaller for the 0.7\% and 0.4\% samples (see Fig. 3b). As $B \rightarrow 7$~T, all the linewidths decrease and converge toward a common value of $\sim$8~meV, which is again very close to the linewidth observed in the nonmagnetic CdSe nanocrystals. This trend indicates a narrowing distribution of Stokes shifts in the ensemble, due to a suppression of spin fluctuations as the Mn$^{2+}$ spins are forced to align by $B$. Note this behavior is also in accord with the fluctuation-dissipation theorem, which mandates that magnetization (spin) fluctuations scale with $\chi$, the magnetic susceptibility: $\langle M^2 \rangle /k_B T \propto \chi$.  For paramagnetic ions, $\chi = \frac{\partial M}{\partial B}$ is simply the slope of the Brillouin function (discussed below), and therefore $\chi \rightarrow 0$ as $B \rightarrow \infty$.  However, we emphasize that the measured linewidths are surprisingly large at zero field -- comparable to $\Delta E$ itself.  Given that $\Delta E$ should scale approximately as $N$ (the average number of Mn$^{2+}$ per nanocrystal), then the broad linewidths at $B$=0 are in contrast with simple expectations that $\Gamma$ should scale as $\sqrt{N}$.  Furthermore, the resonant PL peaks are slightly asymmetric -- the distribution's tail is longer on the low-energy side (most clearly seen at large $B$, where the phonon replicas do not overlap). Again this contrasts with the expectation of symmetric distributions that derives from stochastic (gaussian) $\sqrt{N}$ fluctuations alone. As discussed below, the unexpectedly broad widths and asymmetric lineshapes are due to the statistics of the overlap of randomly-placed Mn$^{2+}$ ions with $\psi_h(\textbf{r})$, the hole's spatially nonuniform envelope wavefunction in the nanocrystal.

To estimate $B_{\rm{ex}}$ and the degree of Mn$^{2+}$ spin alignment due to EMP formation, we must compare the measured polaron binding energy ($\Delta E - 5$~meV) to the giant Zeeman splitting of the 1S exciton absorption that occurs when the Mn$^{2+}$ spins are intentionally aligned by applied fields $B$. Figure 3c shows the total Zeeman splitting, $E_\textrm{Z}$, for the 1.6\% and 0.7\% CdMnSe nanocrystals as determined from polarized absorption spectroscopy (the inset shows the right- and left-circularly polarized absorption spectra of the 1S exciton at 6~T and 3~K).

As expected for Mn$^{2+}$-doped nanocrystals \cite{NorrisNL, BeaulacAM, Bussian}, $E_\textrm{Z}$ is very large and follows a modified Brillouin form:
\begin{equation}
E_\textrm{Z}(B,T) = g_{x} \mu_\textrm{B} B + E_{spd} \langle S_z \rangle \approx E_{spd} \mathcal{B}_{\frac{5}{2}} \left( \frac{5g_{\rm Mn}\mu_{\rm B} B}{2 k_{\rm B} (T+T_0)} \right ).
\end{equation}
The first term describes the small Zeeman splitting that is due to the intrinsic $g$-factor of excitons in CdSe nanocrystals ($\left| g_{x}\right| \simeq 1.4$). This term is very small ($\sim$0.5 meV at 6~T) and can be ignored. The second term accounts for the additional exciton splitting due to \textit{sp-d} exchange interactions with the Mn$^{2+}$ ions, which have average spin projection $\langle S_z \rangle$ along $B$. $\langle S_z \rangle$ follows a Brillouin function $\mathcal{B}_{\frac{5}{2}}$, which describes the field- and temperature-dependent paramagnetism of the spin-5/2 Mn$^{2+}$ ions. Here, $g_{\rm Mn}$=2 is the Mn$^{2+}$ \textit{g}-factor, $\mu_\textrm{B}$ is the Bohr magneton, and $k_\textrm{B}$ is the Boltzmann constant. The saturated magnitude of this second term, $E_{spd}$, is sample-specific and depends on the net overlap of $| \psi_{e,h}(\textbf{r})| ^2$, the probability density of the carrier envelope wave functions, with the embedded Mn$^{2+}$ ions.  In bulk diluted magnetic semiconductors, $E_{spd}$ is typically given within the mean-field and virtual crystal approximation as $E_{spd}=x S_\textrm{eff} (N_0 \alpha - N_0 \beta)$, where $x$ is the Mn concentration, $S_\textrm{eff}$ is the effective Mn$^{2+}$ spin ($\approx 5/2$ for small $x$), and $N_0 \alpha$ and $N_0 \beta$ are the exchange constants characterizing the \textit{s-d} and \textit{p-d} interactions for electrons and holes, respectively.

\begin{figure} [tbp]
\includegraphics [width=0.9 \columnwidth] {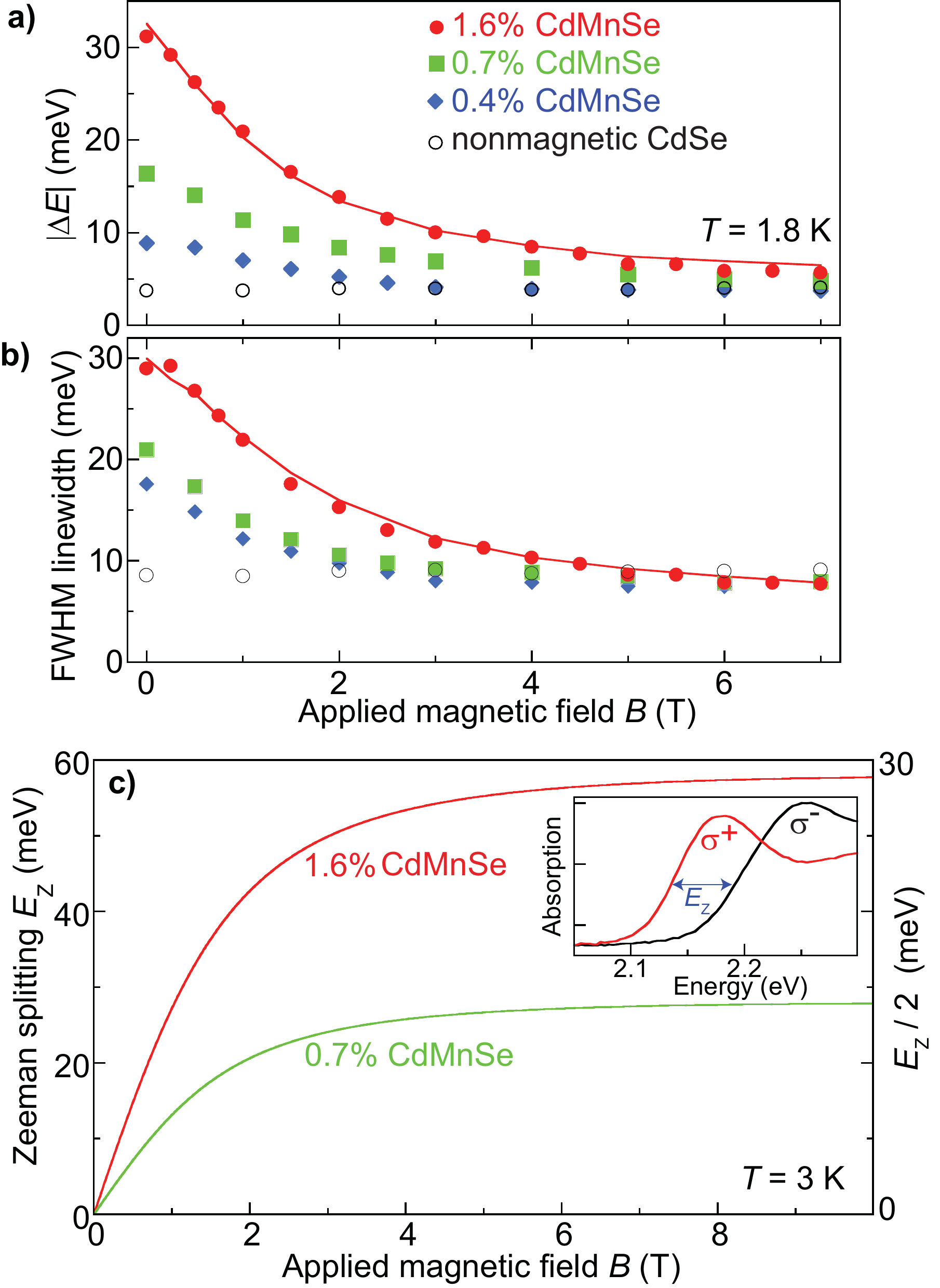}
\caption{(a,b) The measured Stokes shift ($\Delta E$) and linewidth ($\Gamma$) of the resonant PL peaks, for all samples at 1.8~K. At $B$=0, $\Delta E$ and $\Gamma$ are largest in the most heavily doped nanocrystals. As $B \rightarrow 7$~T, both $\Delta E$ and $\Gamma$ decrease and converge toward values observed in the nonmagnetic CdSe nanocrystals.  This indicates a suppression of EMP binding energy and reduction of Mn$^{2+}$ spin fluctuations, respectively. The solid red lines show calculated $\Delta E$ and $\Gamma$ for the 1.6\% CdMnSe nanocrystals, from Monte Carlo simulations. (c) Brillouin functions corresponding to the measured total Zeeman splitting $E_\textrm{Z}$ of the 1S exciton absorption peak in the 1.6\% and 0.7\% CdMnSe nanocrystals at 3~K (see Eq. 1).  $E_\textrm{Z}$ was measured by circularly-polarized absorption; the inset shows example spectra at $B$=6~T and $T$=3~K. $E_\textrm{Z}$ saturates at $E_{spd}$=58~meV and 28~meV, respectively, in these samples. The polaron binding energy can be directly compared with $\frac{1}{2} E_\textrm{Z}$ to determine the degree of Mn$^{2+}$ spin alignment due to EMP formation.}
\label{Fig3}
\end{figure}

In large $B$ where the Mn$^{2+}$ spins are completely aligned, we find that $E_\textrm{Z}$ saturates at $E_{spd} \approx $~28~meV and 58~meV in the 0.7\% and 1.6\% CdMnSe nanocrystals, respectively. Therefore, if an exciton at $B$=0 forms an EMP and completely aligns all the Mn$^{2+}$ spins within its wavefunction envelope, then these aligned spins acting back on the exciton will lower its energy by $\frac{1}{2}E_{spd}$ on average. Comparing the EMP binding energy with $\frac{1}{2}E_{spd}$ therefore provides a means to roughly estimate both the degree of Mn$^{2+}$ spin alignment due to EMP formation, and also the exchange field $B_{\rm{ex}}$ that the exciton exerts on the Mn$^{2+}$ spins.

In the 0.7\% CdMnSe nanocrystals, the measured EMP binding energy of 12~meV at 1.8~K is close to $\frac{1}{2}E_{spd}$ (=14~meV), indicating that the Mn$^{2+}$ ions in these nanocrystals are nearly completely aligned by EMP formation at $B$=0.  Similarly, the 26~meV EMP binding energy measured in the 1.6\% CdMnSe nanocrystals is close to $\frac{1}{2}E_{spd}=29$~meV, again indicating nearly complete alignment of the Mn$^{2+}$ ions due to EMP formation at 1.8~K. In principle, the exchange field $B_{\rm{ex}}$ can be equated with the magnetic field required to achieve this degree of alignment.  In practice, however, such estimates are accurate only when the degree of Mn alignment is modest and the polaron binding energy is much less than $\frac{1}{2}E_{spd}$, as shown in prior works \cite{YakovlevChapter}.  Here at 1.8~K, when the EMP binding energy approaches its maximum value of $\frac{1}{2}E_{spd}$ and the Mn alignment saturates, estimates of $B_{\rm{ex}}$ become very sensitive to any small systematic offsets, and we can infer only that $B_{\rm{ex}}$ is large, at least of order several tesla but possibly much larger.

Temperature-dependent studies provide a more accurate estimate of $B_{\rm{ex}}$. Figure 4a shows resonant PL spectra from the 1.6\% CdMnSe nanocrystals from 2.5-20~K, at $B$=0. In qualitative agreement with prior studies of CdMnTe/CdMgTe superlattices \cite{Mackh1}, the resonant Stokes shift $\Delta E$ decreases as temperature increases.  This indicates a reduction of the EMP binding energy, because the Brillouin-like magnetization of the Mn$^{2+}$ spins, $\langle S_z \rangle$, no longer saturates in the exchange field $B_{\rm{ex}}$ at elevated temperatures (put differently, the Mn$^{2+}$ susceptibility decreases as temperature rises).  This can also be understood by explicitly considering the difference between final and initial exciton energies given by the $B$- and $T$-dependent Zeeman energy from Eq. (1): The quantity $\Delta E \approx \frac{1}{2} [E_\textrm{Z}(B=B_\textrm{ex}) - E_\textrm{Z}(B=0)]$ decreases as temperature rises.

\begin{figure} [tbp]
\includegraphics [width=0.9 \columnwidth] {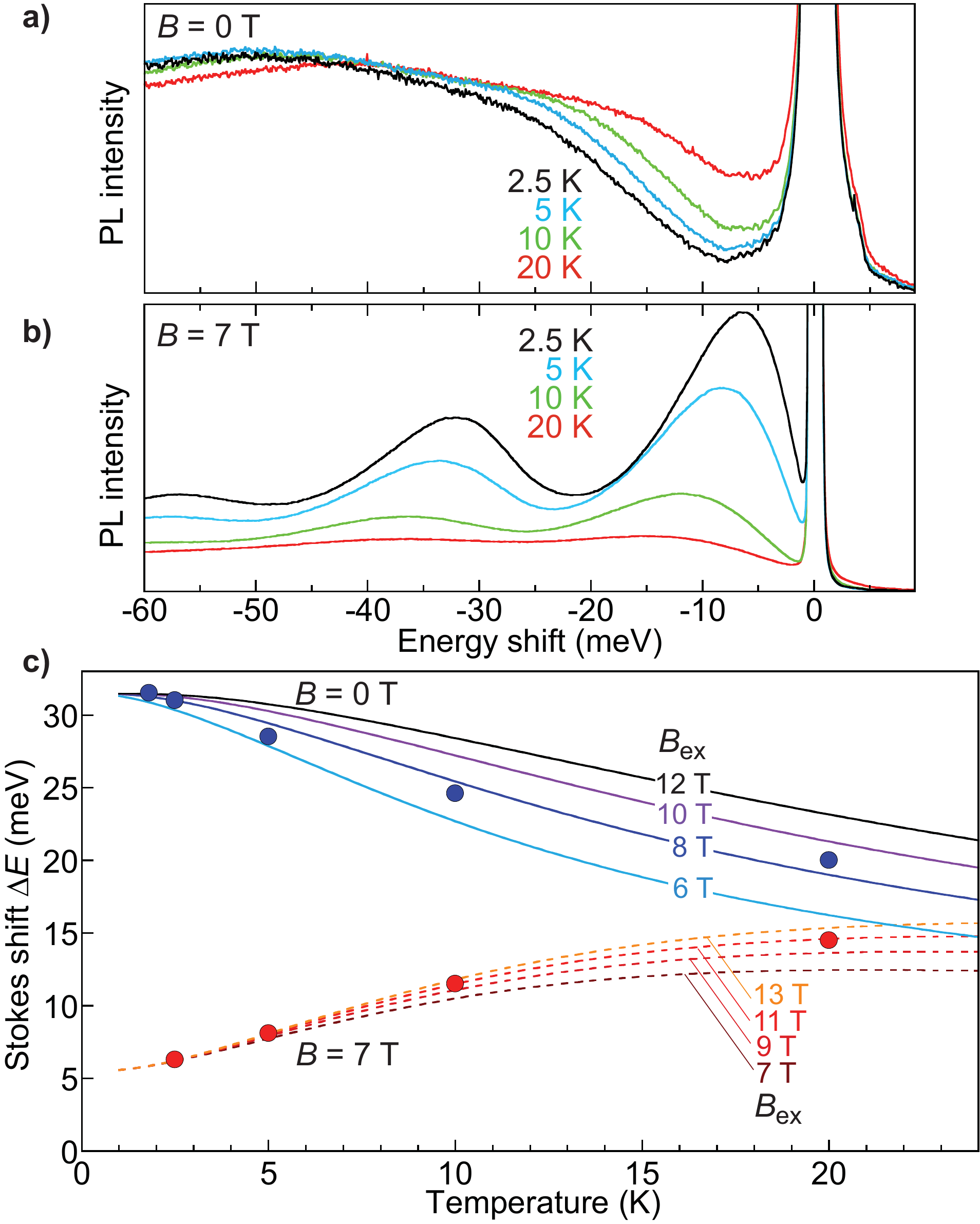}
\caption{(a) Temperature-dependent resonant PL spectra from the 1.6\% CdMnSe nanocrystals at $B$=0~T. (b) Same, but at $B$=7~T. Note that the spectra shift in the opposite direction (see text). (c) Comparing the measured resonant Stokes shifts $\Delta E$ (points) with calculations based on the known field- and temperature-dependent exciton Zeeman splitting for these nanocrystals (the experimentally-observed 5~meV offset has been added to the calculations). Reasonable agreement with the data at 0~T and 7~T are found using an exchange field $B_\textrm{ex}$= 8~T and 11~T, respectively. }
\label{Fig4}
\end{figure}

Similarly, Fig. 4b shows temperature-dependent resonant PL spectra at $B$=7~T. In contrast to the zero-field case, here $\Delta E$ \textit{increases} with increasing temperature. This can also be understood by considering final and initial exciton energies: The quantity $\Delta E \approx \frac{1}{2} [E_\textrm{Z}(B=7 \textrm{T} + B_\textrm{ex}) - E_\textrm{Z}(B=7 \textrm{T})]$ is small at low temperatures (the Mn$^{2+}$ spins are already initially saturated and the additional $B_\textrm{ex}$ has little effect), but \textit{increases} as temperature rises (the Mn$^{2+}$ are no longer initially saturated, and can be further aligned by $B_\textrm{ex}$).

Figure 4c compares the measured $\Delta E$ (points) with expectations from the Brillouin function (lines) for both cases.  The best agreement is found using $B_\textrm{ex} \approx$ 8~T for the case of zero applied field, and, in reasonable concurrence, $B_\textrm{ex} \approx$ 11~T when $B$=7~T. Thus, we conclude that $B_{\rm{ex}}$ lies in the range of 8-11~T for the 1.6\% CdMnSe nanocrystals. Similar values of $B_{\rm{ex}}$ are obtained for the other samples. This large value of the exchange field $B_{\rm{ex}}$ exceeds typical values observed in epitaxially-grown and lightly-doped superlattices and quantum dots ($\sim$3~T) \cite{Kavokin, Bacher}, but agrees rather well with straightforward theoretical estimates of $B_{\rm{ex}}$ for colloidal nanocrystals of this size, as discussed immediately below.

First, we adopt the common assumption that $B_{\rm{ex}}$ derives entirely from the spin of the hole, and neglect the influence of the electron spin \cite{YakovlevChapter, Maksimov, Kavokin}. This assumption is justified because the \textit{p-d} exchange constant $| N_0 \beta |$ significantly exceeds the \textit{s-d} exchange constant $| N_0 \alpha |$: in bulk CdMnSe, $| N_0 \beta |$ = 1.27~eV while $| N_0 \alpha |$=0.23~eV \cite{DMSbook}.  Moreover, $N_0 \alpha$ is likely even further reduced in colloidal nanostructures due to quantum confinement effects \cite{Bussian, Yu, Fainblat, Merkulov1999}. Next, we recall that $B_{\rm{ex}}$ is actually position-dependent within a nanocrystal and scales with the probability amplitude of the hole's envelope wavefunction \cite{YakovlevChapter, Kavokin}:
\begin{equation}
B_{\rm{ex}}(\textbf{r}) = \frac{1}{3 \mu_\textrm{B} g_\textrm{Mn}} \beta J | \psi_h(\textbf{r})|^2,
\end{equation}
where $J=\frac{3}{2}$ is the hole spin. To simplify Eq. (2), we use the popular ``exchange box" model \cite{Dorozhkin, Kavokin, YakovlevChapter} wherein $| \psi_h(\textbf{r})|^2$ and therefore $B_{\rm{ex}}(\textbf{r})$ are taken to be constant within an effective hole localization volume $V$, and zero everywhere else. In this model, all Mn$^{2+}$ ions within $V$ interact equally strongly with the hole. We then obtain the following relationship between $B_{\rm{ex}}$, the \textit{p-d} exchange constant, and polaron volume:
\begin{equation}
B_{\rm{ex}} = \frac{| N_0 \beta |}{2 \mu_\textrm{B} g_\textrm{Mn}} \frac{1}{N_0 V},
\end{equation}
where $N_0$ is the number of cations per unit volume ($N_0 \approx 18.0$/nm$^3$ for CdSe). $B_{\rm{ex}}$ is therefore very sensitive to $V$, which can be significantly less than the physical volume of the nanocrystal because $\psi_h(\textbf{r})$ approaches zero towards the nanocrystal surface. In general, $V$ can be defined \cite{Kavokin, Kusrayev} as $V\equiv (\int| \psi_h(\textbf{r}) |^4~d^3\textbf{r})^{-1}$. If the nanocrystals are modeled as infinite spherical potential wells with radius $a$, then the 1S carrier envelope wavefunctions have functional form $\psi(\textbf{r}) \propto \textrm{sinc}(\pi r/a)$, and $V$ is only $\sim$36\% of the total nanocrystal volume [here, $\textrm{sinc}(x) = \textrm{sin}(x)/x$]. Our nanocrystals have 5~nm average diameter, from which we estimate within this ``box" approximation that $B_{\rm{ex}} \approx 13$~T, which is in reasonable agreement with the values of 8-11~T that were experimentally determined from resonant PL as shown in Fig. 4. Variations in the exact spatial extent and functional form of $\psi_h(\textbf{r})$ will also influence $V$ and therefore $B_{\rm{ex}}$.

Per Eq. 3, $B_{\textrm{ex}}$ scales inversely with nanocrystal volume, and therefore small changes in nanocrystal diameter will significantly influence $B_{\textrm{ex}}$.  However, the measured resonant Stokes shift depends only on $E_{spd}$ (assuming complete EMP formation) and therefore only on the effective Mn concentration in the nanocrystal (see Eq. 1), and is therefore in principle independent of nanocrystal size.

We note that some earlier studies have analyzed the temperature dependence of conventional (\textit{non}-resonant) PL to infer the properties of EMPs. For example, in Ref. \cite{BeaulacScience}, very large values of $B_{\textrm{ex}}$ in the range of 75-120~T were inferred in CdMnSe nanocrystals with 4.3-5.0~nm diameters. A similar analysis of conventional PL from our 1.6\% CdMnSe nanocrystals suggests an exchange field of order 40~T (see Supporting Information), which is 4-5 times larger than the value of $B_{\textrm{ex}}$ determined by our resonant PL studies, and 3 times larger than theoretical expectations. Analysis of EMP properties based on conventional PL/absorption spectroscopy may therefore also be influenced by effects due to inhomogeneous broadening in the ensemble and/or temperature-dependent energy relaxation within the broad manifold of states that comprise the exciton fine structure \cite{NorrisPRB1996}. Both effects are significantly mitigated in resonant PL studies.

\begin{figure} [tbp]
\includegraphics [width=0.8 \columnwidth] {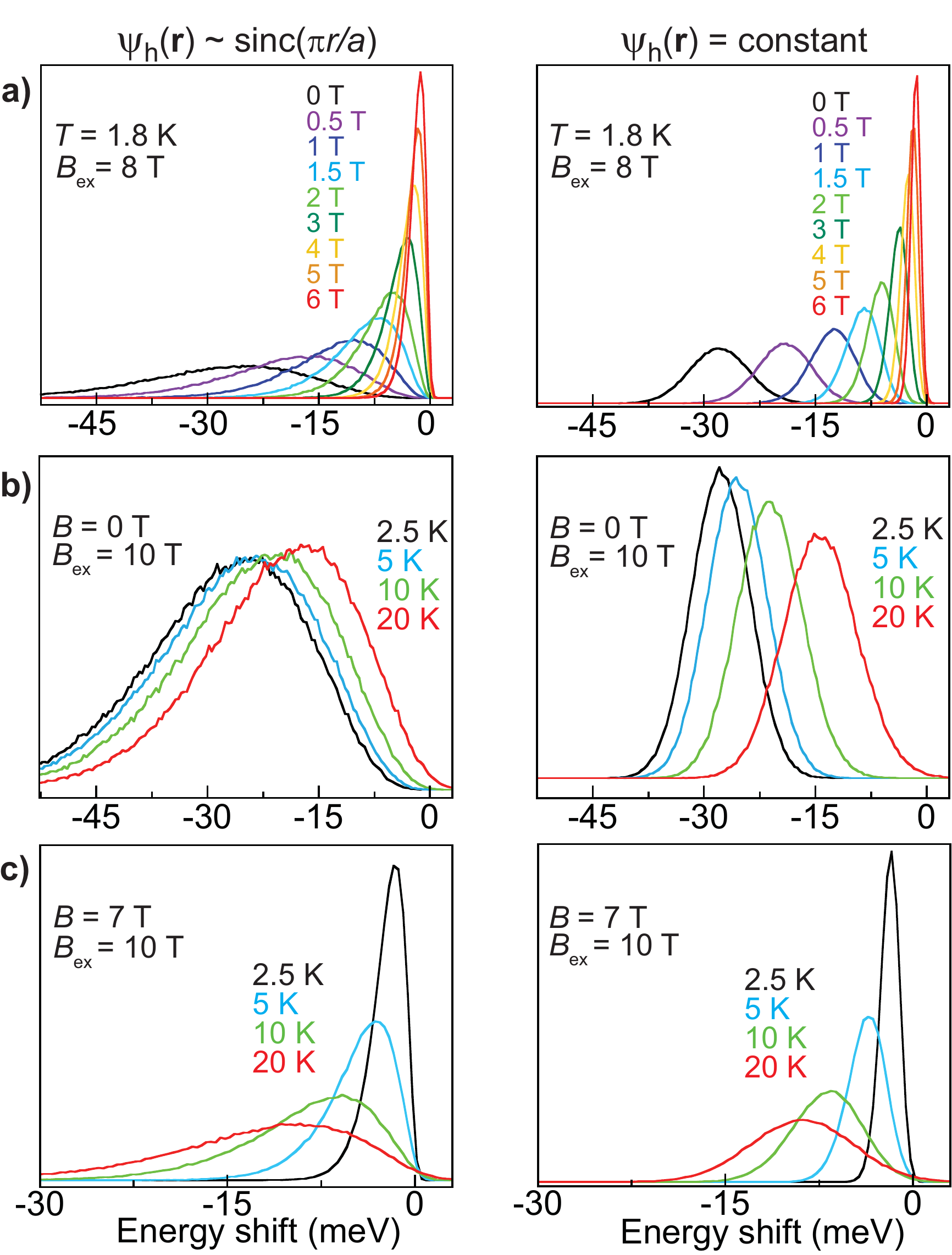}
\caption{Numerical Monte Carlo simulations of the resonant PL spectra.  For clarity, optical phonon replicas are not included. Here, $\langle N \rangle = 20$ (corresponding to the 1.6\% CdMnSe nanocrystals). Spectra are calculated using both $\psi_h(\textbf{r}) \propto \textrm{sinc}(\pi r/a)$ and $\psi_h(\textbf{r})$=constant (left and right panels, respectively). The additional 5~meV offset of the resonant Stokes shift that is observed experimentally is not included in these simulations. (a) Calculated spectra at $T$=1.8~K for increasing applied field $B$ (compare with experimental data in Fig. 2).  (b,c) Calculated spectra at $B$=0 and $B$=7~T for increasing temperature (compare with data in Figs. 4a,b). The broader linewidths and asymmetric line shapes are accurately reproduced when using a realistic functional form of $\psi_h(\textbf{r})$, but are not reproduced within a simple ``box" model where $\psi_h(\textbf{r})$ is constant.}
\label{Fig5}
\end{figure}

Finally, to quantitatively interpret the unexpectedly large linewidths and the asymmetry of the resonant PL peaks that we observed experimentally, we model these resonant PL spectra using a simple numerical Monte Carlo method (see Supporting Information). These simulations account for the effects of fluctuating Mn$^{2+}$ spins and their random spatial distribution within the nanocrystals. Figure 5 shows calculated resonant PL spectra with increasing field and temperature. Phonon replicas are ignored for clarity. The pronounced shift and narrowing of the spectra with increasing applied field $B$ are captured very well (\textit{e.g.}, compare Fig. 5a with the field-dependent experimental data shown in Fig. 2). Similarly, the temperature-dependent decrease and increase of $\Delta E$ at $B$=0 and $B$=7~T, respectively, are also modeled accurately (\textit{e.g.}, compare Figs. 5b,c with the temperature-dependent data shown in Figs. 4a,b).

Importantly, these simulations allow us to study how the linewidths and the detailed lineshapes of the resonant PL features depend on the exact functional form of the hole wavefunction $\psi_h(\textbf{r})$. For example, Fig. 5 shows results for both the simplest ``box" model wherein $\psi_h(\textbf{r})$ is constant throughout the entire nanocrystal, and also for the more realistic scenario where $\psi_h(\textbf{r}) \propto \textrm{sinc}(\pi r/a)$.  When $\psi_h(\textbf{r})$ is constant, the calculated resonant PL spectra are relatively narrow and symmetric, as expected because all Mn$^{2+}$ spins interact equally strongly with the hole. In this case the random spatial distribution of Mn$^{2+}$ within the nanocrystals is irrelevant: At $B$=0 the resonant Stokes shift is determined by $N$ and the linewidth is determined by the $\sqrt{N}$ spin fluctuations alone.

In contrast, both the much larger linewidth as well as the asymmetric lineshape of the resonant PL peaks are more accurately captured when using $\psi_h(\textbf{r}) \propto \textrm{sinc}(\pi r/a)$. This is because the random spatial distribution of the Mn$^{2+}$ ions now matters: different nanocrystals having the same number of randomly placed Mn$^{2+}$ ions can exhibit very different Stokes shifts depending on whether the Mn are statistically located closer to or further from the center of the nanocrystal.  Mn$^{2+}$ spins located near the nanocrystal center where $| \psi_h(\textbf{r}) |^2$ is large are less probable, but significantly impact the exciton energy. These effects further broaden the measured linewidth, and cause the asymmetric line shape. For direct comparison with experimental data, results from these Monte Carlo simulations are shown for the 1.6\% CdMnSe nanocrystals by the solid red lines in Fig. 3a,b. When the field- and temperature-independent offsets for both $\Delta E$ and $\Gamma$ are included, quite good agreement with the experimental data is obtained.

In summary, resonant PL is demonstrated to be a powerful technique for studying magnetic polarons in magnetically-doped colloidal nanocrystals. Via a systematic analysis of the Stokes shifts and linewidths of resonant PL spectra, the binding energies of EMPs and Mn$^{2+}$ spin fluctuation properties are directly measured as a function of temperature, applied magnetic field, and magnetic doping concentration. In these strongly quantum confined CdMnSe nanocrystals, the exchange field $B_\textrm{ex}$ that is `seen' by the Mn$^{2+}$ due to an exciton is determined to be approximately 10~T, in very reasonable agreement with theoretical expectations for nanocrystals of this size. These studies highlight the utility of resonant PL as an important tool for studies of collective phenomena in new colloidal nanomaterials.

We thank N. Sinitsyn for useful discussions and T. A. Baker for help with film preparation.  W.L., V.I.K., and S.A.C. were supported by the Office of Chemical Sciences, Biosciences, and Geosciences of the DOE Office of Basic Energy Sciences. W.D.R. was supported by the Los Alamos LDRD program. D.R.Y. acknowledges support of the Deutsche Forschungsgemeinschaft via ICRC TRR 160. All optical measurements were performed at the National High Magnetic Field Laboratory, which is supported by NSF DMR-1157490.

\newpage

\section{Supplemental Information}

\subsection{Nanocrystal Synthesis, Film Preparation, and Experimental Setup}

Wurtzite CdSe nanocrystals, 5 nm in diameter and capped with octadecylphosphonic acid (ODPA) were synthesized following Ref.~\cite{Carbone}.  Before Mn$^{2+}$ doping, the ODPA ligands were partially replaced with oleic acid. The preparation of lightly-doped Cd$_{1-x}$Mn$_x$Se nanocrystals followed published procedures with necessary modifications \cite{Vlaskin}. The Mn$^{2+}$ concentration was determined from the giant Zeeman splitting of the 1S exciton absorption peak, measured at low temperatures and in applied magnetic fields $B$ to 7~T.

Optical-quality films exhibiting minimal scattering were essential for these resonant photoluminescence (PL) experiments. Typically, 2\% weight-to-weight (w/w) polyvinylpyrrolidone (PVP) in CHCl$_3$ was mixed with a 4~mL nanocrystal/CHCl$_3$ solution.  After centrifuging this mixture, the pellet was then re-dissolved in 600~$\mu$L of 1\% PVP in 20\% w/w butanol/chloroform.  Drops of this reformed mixture were spread onto glass cover slips and then spun at 500~rpm for 120~seconds followed by an additional 60~seconds at 2000~rpm.  This procedure produced thick ($\sim$20~$\mu$m) homogenous nanocrystal films. Ref. ~\cite{Rice} contains additional details about the nanocrystals and film preparation.

All measurements were performed with the samples mounted in the variable-temperature insert of a 7~T magneto-optical cryostat. Magnetic fields $B$ were applied normal to the nanocrystal film and parallel to the direction of optical excitation/detection -- \textit{i.e.}, the Faraday geometry. For resonant PL studies, a low-power continuous-wave dye laser with a narrow linewidth ($\sim$40 GHz) was tuned to the far red (low-energy) side of the 1S exciton absorption resonance in order to excite only a narrow distribution of the lowest-energy 1S excitons in the ensemble and to avoid effects due to energy transfer or photon re-absorption. The resonant PL was detected by a 500~mm spectrometer and a LN$_2$-cooled CCD. We confirmed, at both $B$=0 and 7~T, that the measured resonant Stokes shifts were essentially independent of the exact photon energy of the pump laser, provided that it was tuned well below the 1S exciton absorption peak (in accord with prior resonant PL studies of magnetic polarons in quantum wells and epilayers \cite{Mackh}). For all the data shown, we pumped and detected circularly polarized light. No evidence of optical orientation of magnetic polarons \cite{Yakovlev} was ever observed, likely due to the random nanocrystal orientation.

\subsection{Conventional (Non-resonant) Magneto-PL Measurements}

\begin{figure} [h]
\includegraphics [width=0.98 \columnwidth] {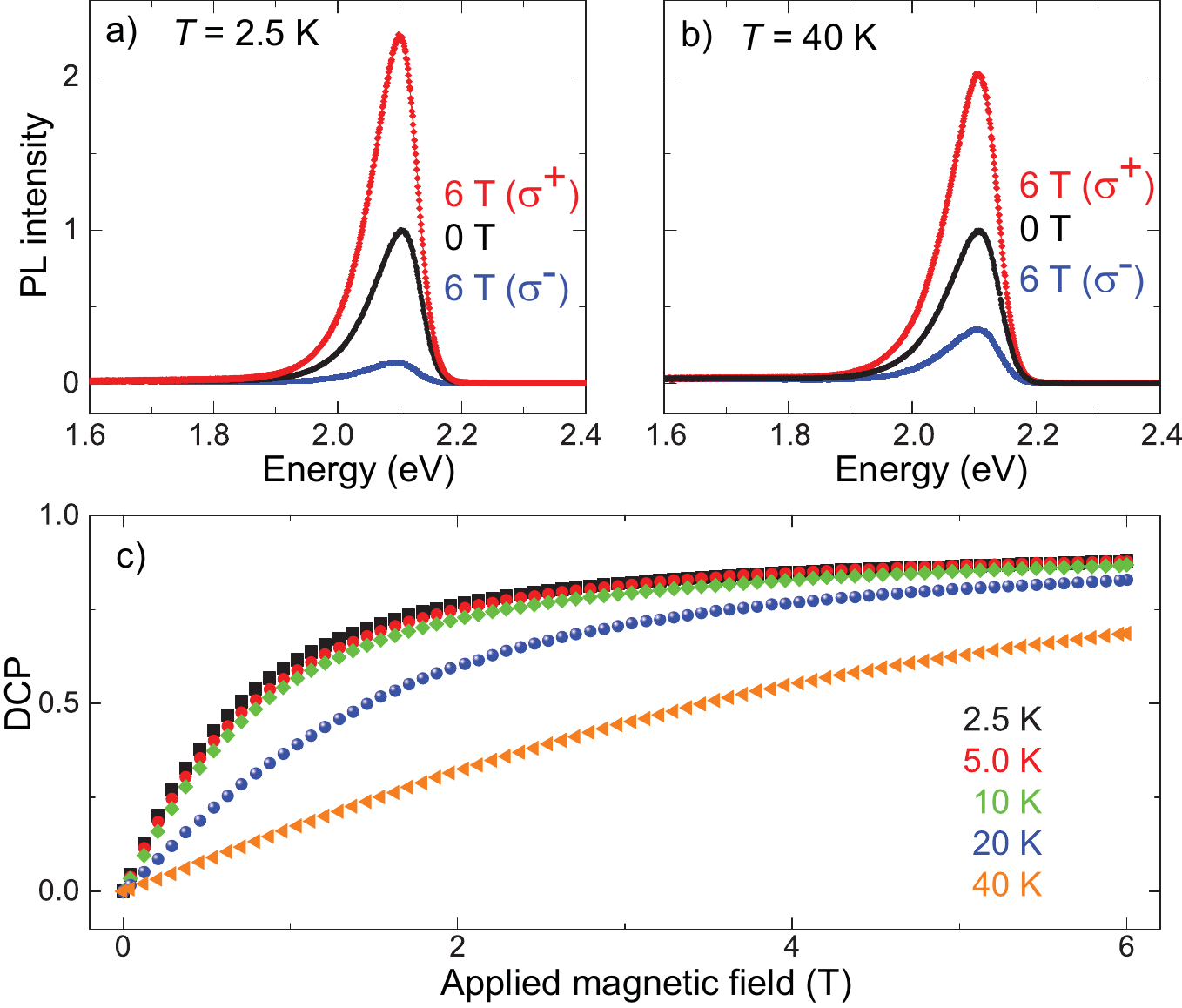}
\caption{(a,b) Non-resonant circularly polarized PL from the 1.6\% CdMnSe nanocrystals at 2.5~K and at 40~K. (c) The degree of circular polarization (DCP) vs. $B$, at different temperatures.}
\label{MCPL_fig}
\end{figure}

Conventional (non-resonant) polarization-resolved PL  was measured on all samples as a function of temperature and magnetic field. High-energy excitons were excited by a weak, linearly-polarized, 405~nm (3.06~eV) cw diode laser. Magneto-PL from the 1.6\% CdMnSe nanocrystals are shown in Fig. \ref{MCPL_fig}. The PL develops a significant degree of circular polarization $\left({\rm DCP} = \frac{I_{\sigma +} - I_{\sigma -}}{I_{\sigma +} + I_{\sigma -}}\right)$ with increasing $B$. The circularly-polarized PL peaks at $B$=6~T are only very slightly shifted from the 0~T PL peak (by $<5$~meV, which is much less than the $\sim$30~meV EMP binding energy), consistent with a scenario in which excitons recombine from a fully-formed low-energy EMP state even at $B$=0.  At the lowest temperatures, the DCP saturates at $\sim$0.85 by 3~T (in comparison, nonmagnetic CdSe nanocrystals require larger applied magnetic fields to achieve similar DCP saturation \cite{FurisJPC, Zeke}). However, the DCP actually saturates much more gradually with $B$ than is expected from a simple model of excitons thermally populating two spin states that are split by the giant Zeeman splitting $E_{\rm Z} (B,T)$ in these nanocrystals. As discussed previously \cite{Nelson}, aspects of this behavior in conventional PL measurements may be due to the details of EMP formation, the influence of Mn$^{2+}$ spin fluctuations, and the random nanocrystal orientation.

\subsection{Time-Resolved PL Measurements}

Time-correlated single-photon counting studies were performed using a 405~nm pulsed diode laser ($\sim$70~ps pulse width) and a fast multichannel plate photomultiplier tube attached to a 500~mm spectrometer. This allowed both time- and spectrally-resolved PL decays, from which dynamic redshifts of the emission spectrum were determined.

\begin{figure} [htbp]
\includegraphics [width=0.98 \columnwidth] {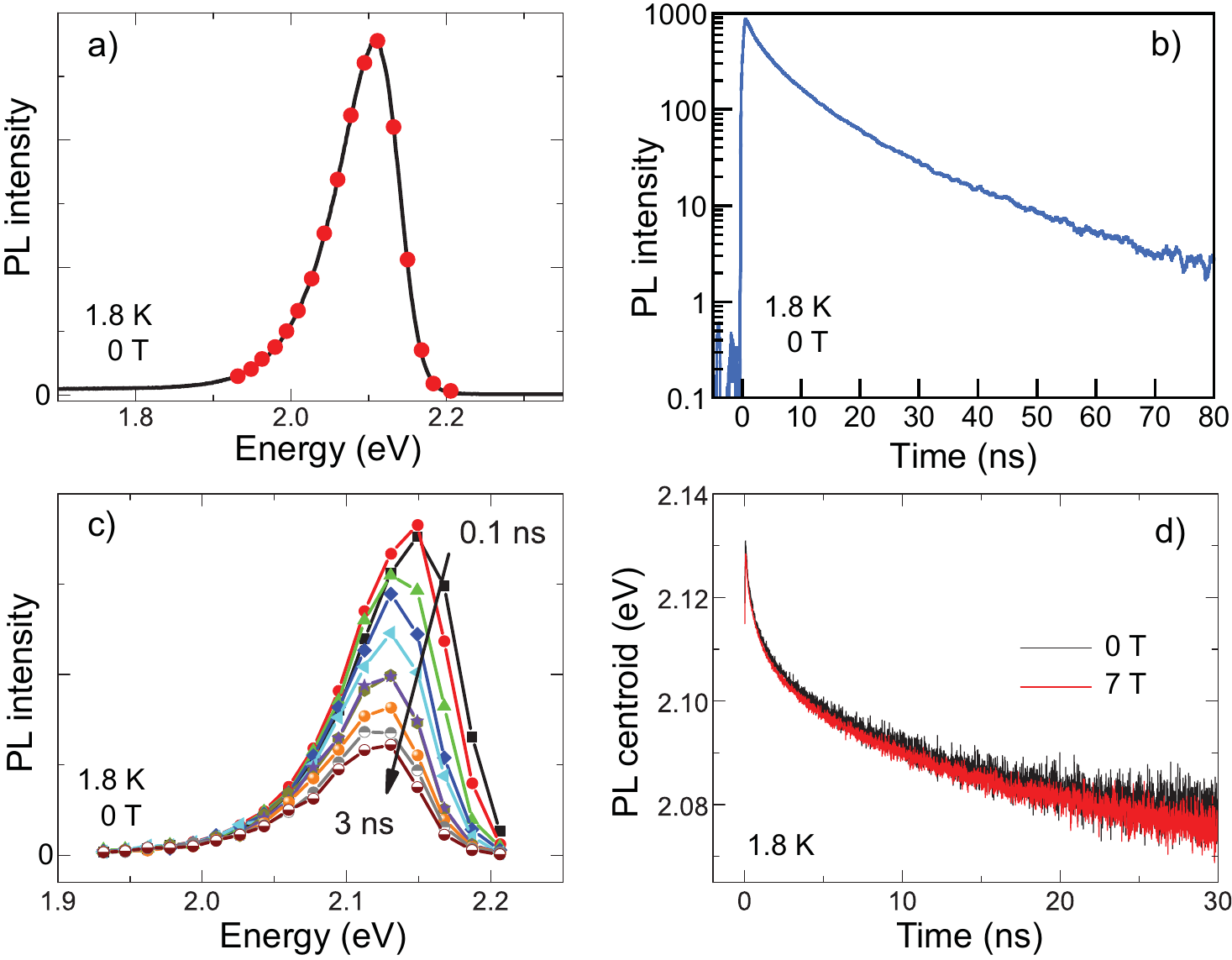}
\caption{(a) Conventional (nonresonant) PL spectrum from the 1.6\% CdMnSe nanocrystals.  The red dots show the specific photon energies at which time-resolved PL was collected. (b) Time-resolved PL acquired at the PL maximum (2.1~eV) exhibits a multi-exponential decay, with the slow component of the decay significantly exceeding 10~ns. These long exciton lifetimes are much longer than the typically sub-ns timescales of EMP formation. (c) ``Instantaneous" PL spectra from 0.1 ns to 3 ns, showing a dynamic redshift. (d) The centroid of the PL emission versus time, showing that the dynamic redshift slows with time, but still continues out past 30~ns, and is very similar at both $B$=0 and $B$=7~T.}
\label{TRPL_fig}
\end{figure}

Figure \ref{TRPL_fig}a shows the conventional (non-resonant) PL spectrum from the 1.6\%~CdMnSe nanocrystals at 1.8~K and 0~T. The red dots show the specific photon energies where time-resolved PL decays were measured. Figure \ref{TRPL_fig}b shows the PL decay measured at the PL maximum (2.1~eV). The decay is multiexponential with a slow component of order 20~ns, which is over an order of magnitude longer than the (sub-nanosecond) timescales typically required for complete EMP formation \cite{Nelson, Yakovlev}. Thus, EMP formation proceeds to an equilibrium condition and is not interrupted by exciton recombination. 

Note that the 20~ns lifetime, although sufficient for complete EMP formation, is still much shorter than the $\sim$1000~ns PL lifetimes typically observed at low temperatures from \emph{non}-magnetic CdSe nanocrystals \cite{CrookerAPL} where the ground state exciton is known to be an optically-forbidden $J = \pm 2$ ``dark" exciton. PL lifetimes in the range of 10-20~ns are typical for optically-allowed $J = \pm 1$ ``bright" excitons. It is tempting, therefore, to associate the 20~ns PL lifetime observed in these CdMnSe nanocrystals with emission from an optically-allowed bright exciton ground state. However, as discussed in Fig. 3 of the main text, the resonant PL data clearly show a 5~meV offset in $\Delta E$ at large $B$, which is consistent with a dark exciton ground state in these CdMnSe nanocrystals. To understand the short PL lifetimes in this case, recall that applied magnetic fields will mix dark exciton states with their bright counterparts in randomly-oriented nanocrystal ensembles, leading to significantly accelerated PL dynamics from nominally dark excitons at low temperatures \cite{FurisJPC}. The influence of large \emph{sp-d} exchange interactions in Mn-doped nanocrystals is expected to have a similar effect, even at zero applied field. Further studies are clearly needed in order to pinpoint the nature of the exciton ground state in Mn-doped colloidal nanocrystals.

Figure \ref{TRPL_fig}c shows the reconstructed ``instantaneous" PL spectra within the first 3 ns after nonresonant photoexcitation.  A dynamic redshift is observed on these short timescales. Figure ~\ref{TRPL_fig}d shows the centroid of the PL spectrum on longer timescales, where it is clear that the dynamic redshift slows but nonetheless continues out to very long timescales of order 30~ns. While this dynamic redshift is consistent with EMP formation and subsequent slow reorientation/relaxation of the EMPs \cite{Beaulac, Nelson}, we also note that very similar dynamic redshifts of exciton PL have been observed in ensembles of \textit{non}-magnetic CdSe nanocrystals, due for example to F\"orster energy transfer \cite{CrookerPRL2002, Liu}. The dynamic redshifts observed here are therefore also consistent with energy transfer within the ensemble. Very likely, both effects contribute to the dynamic PL redshift. Interestingly, Figure ~\ref{TRPL_fig}d also shows that there is very little difference between the dynamic redshift observed at $B$=0 and $B$=7~T.

\subsection{Temperature Dependence of Non-resonant PL}

Previous studies \cite{Bacher, Maksimov, Beaulac, Nelson} have used the temperature dependence of the \textit{global} (\textit{i.e.}, the \textit{non}-resonant) Stokes shift -- by which we mean the large energy difference between the 1S exciton absorption peak and the nonresonant PL peak -- to infer the presence of magnetic polarons and to estimate the polaron binding energy and the exchange field $B_{ex}$.  For example, Ref. \cite{Beaulac} utilized this non-resonant technique on CdMnSe colloidal nanocrystals ($4.3 - 5.0$ ~nm diameter) and inferred surprisingly large values of $B_{\textrm{ex}}$ from 75-120~T.

\begin{figure*} [htbp]
\includegraphics [width=5 in] {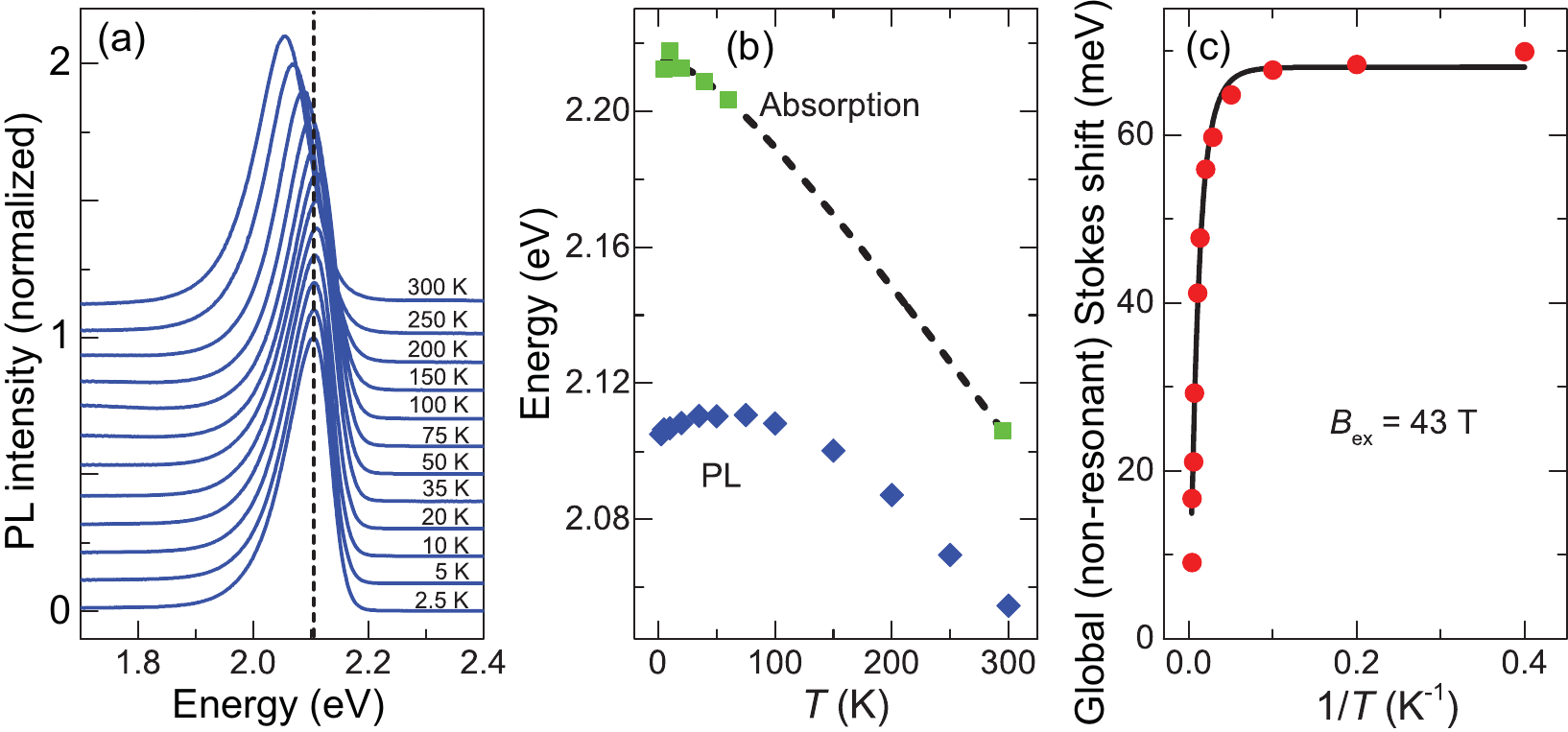}
 \caption{(a) Non-resonant PL spectra from the 1.6\% CdMnSe nanocrystals from $2.5-300$~K.  (b) 1S exciton absorption peak energy (green) and non-resonant PL peak energy (blue).  The dotted line is the predicted absorption peak energy at intermediate temperatures based on fitting to a Varshni model. (c) The global (non-resonant) Stokes shift with a 40~meV offset subtracted.  A Brillouin fit to the data is achieved using $B_{\rm ex} \sim$~43~T, which is a factor of 4-5 times larger than the values of $B_{\rm ex}$ that were directly measured via resonant PL studies (see main text).}
\label{T-dependent_PL_fig}
\end{figure*}

We performed a similar analysis of the global Stokes shift on our 1.6\% CdMnSe nanocrystal sample.  Conventional absorption and (non-resonant) PL were measured from room temperature down to 2.5~K. As shown in Fig.~\ref{T-dependent_PL_fig}, the 1S exciton absorption peak exhibits the usual monotonic Varshni blueshift as the temperature decreases. At intermediate temperatures where the absorption was not explicitly measured, we interpolate the absorption peak position using the Varshni equation (dashed line). The nonresonant PL peak energy also blueshifts upon cooling down to about 50~K, and then redshifts slightly at lower temperatures down to 2.5~K.  This redshift is consistent with an additional lowering of the exciton's final energy by magnetic polaron formation.

The temperature dependence of the non-resonant Stokes shift is shown in Fig.~\ref{T-dependent_PL_fig}c (a constant 40~meV offset has been subtracted from the data). The data are fit to a Brillouin function:  $\mathcal{B}_{5/2}\left(\frac{5g_{\rm Mn}\mu_B B_{\rm ex}}{2k_B (T + T_0)}\right)$. The value of $B_{\rm ex}$ obtained using this protocol is a very large 43~T, which is a factor of 4-5 larger than the value that we directly measured using resonant PL spectroscopy (see main text), and significantly larger than the value of $\sim$13~T predicted by theory. Though the low-temperature redshift of the nonresonant PL spectra is consistent with EMP formation, accurate determination of $B_{\textrm{ex}}$ based strictly on the temperature dependence of global Stokes shifts and/or non-resonant PL measurements may be influenced by additional factors that are unrelated to EMP formation, such as  inhomogeneous broadening in the ensemble and/or temperature-dependent energy relaxation within the broad manifold of states that comprise the exciton fine structure. Both effects are largely mitigated in resonant PL studies.

\subsection{Numerical Monte Carlo Simulations}

We simulated the measured resonant PL spectra using a numerical Monte Carlo approach, which necessarily accounts for fluctuations of Mn$^{2+}$ spin orientation and distribution within the nanocrystal ensemble. The simulations were performed as follows: For each nanocrystal, we first generate $N$ random locations within a unit sphere for the Mn$^{2+}$ ions. $\langle N \rangle$ ranged from 5 to 20 (corresponding to our 0.4\% - 1.6\% CdMnSe nanocrystals) and could be chosen from an appropriate Poisson distribution. A random spin orientation for each Mn$^{2+}$ is then generated, weighted by the appropriate distribution for the given temperature and applied field $B$. The initial energy of the exciton is calculated by summing over the individual exchange energies from each Mn$^{2+}$ ion, weighted by the probability amplitude of the properly normalized hole wavefunction $| \psi_h(\textbf{r})|^2$.  Then, to account for polaron formation, each of the same $N$ Mn$^{2+}$ ions is assigned new random spin orientation drawn from a new distribution that corresponds to the \textit{total} magnetic field $B+B_{\rm{ex}}(\textbf{r}) = B+B_{\rm{ex}} | \psi_h(\textbf{r}) |^2$. Note that this procedure adopts the simplification that $B$ and $B_{\rm{ex}}$ are parallel and can be added as scalars; in actuality $B$ and $B_{\rm{ex}}$ may not be strictly parallel. The final exciton energy is again obtained via the sum over the individual exchange energies from each Mn$^{2+}$ ion, weighted by $| \psi_h(\textbf{r})|^2$. The Stokes shift $\Delta E$ is given by the difference between initial and final exciton energies, and the distribution of $\Delta E$ is obtained by repeating the simulation hundreds of thousands of times.  A related Monte Carlo approach was described in \cite{Kusrayev2} to simulate bound magnetic polarons in a very dilute regime in bulk CdMnTe.


\begin{references}

\bibitem{NorrisNL}Norris, D. J.; Yao, N.; Charnock, F. T.; Kennedy, T. A. \textit{Nano Lett.} \textbf{2001}, \textit{1}, 3-7. 

\bibitem{Mikulec}Mikulec, F. V.; Kuno, M.; Bennati, M.; Hall, D. A.; Griffin, R. G.; Bawendi, M. G. \textit{J. Am. Chem. Soc.} \textbf{2000}, \textit{122}, 2532-2540. 

\bibitem{Hoffman}Hoffman, D. M.; Meyer, B. K.; Ekimov, A. I.; Merkulov, I. A.; Efros, Al. L.; Rosen, M.; Couino, G.; Gacoin, T.; Boilot, J. P. \textit{Solid State. Comm.} \textbf{2000}, \textit{114}, 547-550. 

\bibitem{Erwin}Erwin, S. C.; Zu, L.; Haftel, M. I.; Efros, Al. L.; Kennedy, T. A.; Norris, D. J. \textit{Nature} \textbf{2005}, \textit{436}, 91-94. 

\bibitem{Ithurria}Ithurria, S.; Guyot-Sionnest, P.; Mahler, B.; Dubertret, B. \textit{Phys. Rev. Lett.} \textbf{2007}, \textit{99}, 265501. 

\bibitem{Bhargava}Bhargava, R. N.; Gallagher, D.; Hong, X.; Nurmikko, A. \textit{Phys. Rev. Lett.} \textbf{1994}, \textit{72}, 416-419.

\bibitem{BeaulacAM}Beaulac, R.; Archer, P. I.; Ochsenbein, S. T.; Gamelin, D. R. \textit{Adv. Funct. Mater.} \textbf{2008}, \textit{18}, 3873-3891. 

\bibitem{Bussian}Bussian, D. A.; Crooker, S. A.; Yin, M.; Brynda, M.; Efros, Al. L.; Klimov, V. I. \textit{Nature Mater.} \textbf{2009}, \textit{8}, 35-40. 

\bibitem{FurdynaJAP}Furdyna, J. K. \textit{J. Appl. Phys.} \textbf{1988}, \textit{64}, R29-R64.

\bibitem{Dietl}Dietl, T. in \textit{Handbook on Semiconductors}, edited by T. S. Moss and S. Mahajan (North-Holland, Amsterdam, 1994).

\bibitem{DMSbook}\textit{Introduction to the Physics of Diluted Magnetic Semiconductors} edited by J. Kossut and J. A. Gaj (Springer-Verlag, Berlin, 2010).

\bibitem{Yu}Yu, J. H.; Liu, X.; Kweon, K. E.; Joo, J.; Park J.; Ko, K.-T., Lee, D. W.; Shen, S.; Tivakornsasithorn, K.; Son, J. S.; Park J.-H.; Kim, Y.-W.; Hwang, G. S.; Dobrowolska, M.; Furdyna, J. K.; Hyeon, T. \textit{Nature Mater.} \textbf{2010}, \textit{9}, 47-53. 

\bibitem{Fainblat}Fainblat, R.; Frohleiks, J.; Muckel, F.; Yu, J. H.; Yang, J.; Hyeon, T.; Bacher, G. \textit{Nano Lett.} \textbf{2012}, \textit{12}, 5311-5317. 

\bibitem{Murphy}Murphy J. R.; Delikanli, S.; Scrace, T.; Zhang, P.; Norden, T.; Thomay, T.; Cartwright, A. N.; Demir, H. V.; Petrou, A. \textit{Appl. Phys. Lett.} \textbf{2016}, \textit{108}, 242406. 

\bibitem{Maksimov}Maksimov, A. A.; Bacher, G.; McDonald, A.; Kulakovskii, V. D.; Forchel, A.; Becker, C. R.; Landwehr, G.; Molenkamp, L. W. \textit{Phys. Rev. B }\textbf{2000}, \textit{62}, R7767-R7770. 

\bibitem{Seufert}Seufert, J.; Bacher, G.; Scheibner, M.; Forchel, A.; Lee, S.; Dobrowolska, M.; Furdyna, J. K. \textit{Phys. Rev. Lett. } \textbf{2002}, \textit{88}, 027402. 

\bibitem{Bacher}Bacher, G.; Maksimov, A. A.; Sch\"omig, H.; Kulakovskii, V. D.; Welsch, M. K.; Forchel, A.; Dorozhkin, P. S.; Chernenko, A. V.; Lee, S.; Dobrowolska, M.; Furdyna, J. K. \textit{Phys. Rev. Lett.} \textbf{2002}, \textit{89}, 127201. 

\bibitem{Dorozhkin}Dorozhkin, P. S.; Chernenko, A. V.; Kulakovskii, V. D.; Brichkin, A. S.; Maksimov, A. A.; Schoemig, H.; Bacher, G.; Forchel, A.; Lee, S.; Dobrowolska, M.; Furdyna, J. K. \textit{Phys. Rev. B } \textbf{2003}, \textit{68}, 195313. 

\bibitem{Mackowski}Mackowski, A.; Gurung, T.; Nguyen, T. A.; Jackson, H. E.; Smith, L. M.; Karczewski, G.; Kossut, J. \textit{Appl. Phys. Lett.} \textbf{2004}, \textit{84}, 3337-3339. 

\bibitem{Besombes}Besombes, L.; L\'eger, Y.; Maingault, L.; Ferrand, D.; Mariette, H.; Cibert, J. \textit{Phys. Rev. Lett.} \textbf{2004}, \textit{93}, 207403. 

\bibitem{Wojnar}Wojnar, P.; Suffczy\'nski, J.; Kowalik, K.; Golnik, A.; Karczewski, G.; Kossut, J. \textit{Phys. Rev. B.} \textbf{2007}, \textit{75}, 155301. 

\bibitem{Abolfath}Abolfath, R. M.; Hawrylak, P.; \v Zuti\'c, I. \textit{Phys. Rev. Lett.} \textbf{2007}, \textit{98}, 207203. 

\bibitem{Sellers}Sellers, I. R.; Oszwaldowski, R.; Whiteside, V. R.; Eginligil, M.; Petrou, A.; Zutic, I.; Chou, W. C.; Fan, W. C.; Petukhov, A. G.; Kim, S. J.; Cartwright, A. N.; McCombe, B. D. \textit{Phys. Rev. B.} \textbf{2010}, \textit{82}, 195320. 

\bibitem{Trojnar}Trojnar, A. H.; Korkusi\'nski, M.; Kadantsev, E. S.; Hawrylak, P.; Goryca, M.; Kazimierczuk, T.; Kossacki, P.; Wojnar, P.; Potemski, M. \textit{Phys. Rev. Lett.} \textbf{2011}, \textit{107}, 207403. 

\bibitem{Klopotowski}K\l opotowski, \L.;  Cywi\'nski, \L.; Wojnar, P.; Voliotis, V.; Fronc, K.; Kazimierczuk, T.; Golnik, A.; Ravaro, M.; Grousson, R.; Karczewski, G.; Wojtowicz, T. \textit{Phys. Rev. B} \textbf{2011}, \textit{83}, 081306. 

\bibitem{Kobak}Kobak, J.; Smole\'nski, T.; Goryca, M.; Papaj, M.; Gietka, K.; Bogucki, A.; Koperski, M.; Rousset, J.-G; Suffczy\'nski, J.; Janik, E.; Nawrocki, M.; Golnik, A.; Kossacki, P.; Pacuski, W. \textit{Nature Commun.} \textbf{2014}, \textit{5}:3191. 

\bibitem{Bhattacharjee}Bhattacharjee, A. K.; Benoit \`a la Guillaume, C. \textit{Phys. Rev. B} \textbf{1997}, \textit{55}, 10613-10620. 

\bibitem{BeaulacScience}Beaulac, R.; Schneider, L.; Archer, P. I.; Bacher, G.; Gamelin, D. R. \textit{Science} \textbf{2009}, \textit{325}, 973-976.

\bibitem{Nelson}Nelson, H. D.; Bradshaw, L. R.; Barrows, C. J.; Vlaskin, V. A.; Gamelin, D. R. \textit{ACS Nano} \textbf{2015}, \textit{9}, 11177-11191.

\bibitem{RiceNNANO}Rice, W. D.; Liu, W.; Baker, T. A.; Sinitsyn, N. A.; Klimov, V. I.; Crooker, S. A. \textit{Nat. Nanotechnol.} \textbf{2016}, \textit{11}, 137-142.

\bibitem{YakovlevChapter}Yakovlev, D. R. \& Ossau, W. ``Introduction to the Physics of Diluted Magnetic Semiconductors" (Springer-Verlag, Berlin, 2010). (Eds. Kossut, J. and Gaj, J.) Chapter 7: Magnetic Polarons.

\bibitem{DietlPRL1995}Dietl, T.; Peyla, P.; Grieshaber, W.; Merle d'Aubign\'e, Y. \textit{Phys. Rev. Lett. }\textbf{1995}, \textit{74}, 474-477. 

\bibitem{Kavokin}Kavokin, K. V.; Merkulov, I. A.; Yakovlev, D. R.; Ossau, W.; Landwehr, G. \textit{Phys. Rev. B} \textbf{1999}, \textit{60}, 16499-16505. 

\bibitem{Zayhowski}Zayhowski J. J.; Jagannath, C.; Kershaw, R. N.; Ridgley, D.; Dwight, K.; Wold, A. \textit{Solid State Comm.} \textbf{1985}, \textit{55}, 941-945. 

\bibitem{Itoh}Itoh, T.; Komatsu, E. \textit{J. Lumin.} \textbf{1987}, \textit{38}, 266-268. 

\bibitem{Yakovlev1992}Yakovlev, D. R.; Ossau, W.; Landwehr, G.; Bicknell-Tassius, R. N.; Waag, A.; Schmeusser, S.; Uraltsev, I. N.  \textit{Solid State Comm.} \textbf{1992}, \textit{82}, 29-32. 

\bibitem{Mackh1}Mackh, G.; Ossau, W.; Yakovlev, D. R.; Waag, A.; Landwehr, G.; Hellmann, R.; G\"obel, E. O. \textit{Phys. Rev. B} \textbf{1994}, \textit{49}, 10248-10258. 

\bibitem{Mackh2}Mackh, G.; Hilpert, M.; Yakovlev, D. R.; Ossau, W.; Heinke, H.; Litz, T.; Fischer, F.; Waag, A.; Landwehr, G.; Hellmann, R.; G\"obel, E. O. \textit{Phys. Rev. B} \textbf{1994}, \textit{50}, 14069-14076. 

\bibitem{Takeyama}Takeyama, S.; Adachi, S.; Takagi, Y.; Aguekian, V. F. \textit{Phys. Rev. B} \textbf{1995}, \textit{51}, 4858-4864. 

\bibitem{Oka}Oka, Y.; Shen, J.; Takabayashi, K.; Takahashi, N.; Mitsu, H.; Souma, I.; Pittini, R. \textit{J. Lumin.} \textbf{1999}, \textit{83-84}, 83-89. 

\bibitem{Fiederling}Fiederling, R.; Yakovlev, D. R.; Ossau, W.; Landwehr, G.; Merkulov, I. A.; Kavokin, K. V.; Wojtowicz, T.; Kutrowski, M.; Grasza, K.; Karczewski, G.; Kossut, J. \textit{Phys. Rev. B} \textbf{1998}, \textit{58}, 4785-4792. 

\bibitem{Kusrayev}Kusrayev, Yu. G.; Kavokin, K. V.; Astakhov, G. V.; Ossau, W.; Molenkamp, L. W. \textit{Phys Rev. B} \textbf{2008}, \textit{77}, 085205. 

\bibitem{CrookerPRL2002}Crooker, S. A.; Hollingsworth, J. A.; Tretiak, S.; Klimov, V. I. \emph{Phys. Rev. Lett.} \textbf{2002}, \emph{89}, 186802.

\bibitem{Liu}Liu, F.; Rodina, A. V.; Yakovlev, D. R.; Golovatenko, A. A.; Greilich, A.; Vakhtin, E. D.; Susha, A.; Rogach, A. L.; Kusrayev, Yu. G.; Bayer, M. \emph{Phys. Rev. B} \textbf{2015}, \emph{92}, 125403.

\bibitem{NorrisPRB1996}Norris, D. J.; Efros, Al. L.; Rosen, M.; Bawendi, M. G. \textit{Phys. Rev. B} \textbf{1996}, \textit{53}, 16347-16354. 

\bibitem{Nirmal}Nirmal, M.; Norris, D. J.; Kuno, M.; Bawendi, M. G.; Efros, Al. L.; Rosen, M. \textit{Phys. Rev. Lett.} \textbf{1995}, \textit{75}, 3728-3731. 

\bibitem{Furis}Furis, M.; Barrick, T.; Petruska, M.; Htoon, H.; Klimov, V. I.; Crooker, S. A. \textit{Phys. Rev. B} \textbf{2006}, \textit{73}, 241313. 

\bibitem{Merkulov1999}Merkulov, I. A.; Yakovlev, D. R.; Keller, A.; Ossau, W.; Geurts, J.; Waag, A.; Landwehr, G.; Karczewski, G.; Wojtowicz, T.; Kossut, J. \textit{Phys. Rev. Lett.} \textbf{1999}, \textit{83}, 1431-1434.

\end{references}

\begin{references}

\bibitem{Carbone}L. Carbone, C. Nobile, M. De Giorgi, F. Della Sala, G. Morello, P. Pompa, M. Hytch, E. Snoeck, A. Fiore, I. R. Franchini, M. Nadasan, A. F. Silvestre, L. Chiodo, S. Kudera, R. Cingolani, R. Krahne, L. Manna, Nano Lett. \textbf{7}, 2942-2950 (2007).

\bibitem{Vlaskin}V. A. Vlaskin, C. J. Barrows, C. S. Erickson, D. R. Gamelin, J. Am. Chem. Soc. \textbf{135}, 14380 (2013).

\bibitem{Rice}W. D. Rice, W. Liu, T. A. Baker, N. A. Sinitsyn, V. I. Klimov, and S. A. Crooker, Nature Nanotech. \textbf{11}, 137 (2016).

\bibitem{Mackh} G. Mackh, W. Ossau, D. R. Yakovlev, A. Waag, G. Landwehr, R. Hellmann, E. O. G\"obel, Phys. Rev. B \textbf{49}, 10248 (1994).

\bibitem{Yakovlev}D. R. Yakovlev and W. Ossau, ``Introduction to the Physics of Diluted Magnetic Semiconductors” (Springer, 2010) Chapter 7, Magnetic Polarons.

\bibitem{FurisJPC}M. Furis, J. A. Hollingsworth, V. I. Klimov, S. A. Crooker, J. Phys. Chem. B \textbf{109}, 15332-15338 (2005).

\bibitem{Zeke}E. Johnston-Halperin, D. D. Awschalom, S. A. Crooker, Al. L. Efros, M. Rosen, X. Peng, A. P. Alivisatos, Phys. Rev. B \textbf{63}, 205309 (2001).

\bibitem{Nelson}H. D. Nelson, L. R. Bradshaw, C. J. Barrows, V. A. Vlaskin, D. R. Gamelin, ACS Nano \textbf{9}, 11177 (2015).

\bibitem{FurdynaJAP}J. K. Furdyna, J. Appl. Phys. \textbf{64}, R29-R64 (1988).

\bibitem{CrookerAPL}S. A. Crooker, T. Barrick, J. A. Hollingsworth, V. I. Klimov, Appl. Phys. Lett. \textbf{82}, 2793 (2003).

\bibitem{Beaulac}R. Beaulac, L. Schneider, P. I. Archer, G. Bacher, and D. R. Gamelin, Science \textbf{325}, 973 (2009).

\bibitem{CrookerPRL2002}S. A. Crooker, J. A. Hollingsworth, S. Tretiak, V. I. Klimov, Phys. Rev. Lett. \textbf{89}, 186802 (2002).

\bibitem{Liu}Feng Liu, A. V. Rodina, D. R. Yakovlev, A. A. Golovatenko, A. Greilich, E. D. Vakhtin, A. Susha, A. L. Rogach, Yu. G. Kusrayev, M. Bayer, Phys. Rev. B \textbf{92}, 125403 (2015).

\bibitem{Bacher}G. Bacher, H. Sch\"omig, M. K. Welsch, S. Zaitsev, V. D. Kulakovskii, A. Forchel, S. Lee, M. Dobrowolska, J. K. Furdyna, B. K\"onig, W. Ossau, Appl. Phys. Lett. \textbf{79}, 524 (2001).

\bibitem{Maksimov}A. A. Maksimov, G. Bacher, A. McDonald, V. D. Kulakovskii, A. Forchel, C. R. Becker, G. Landwehr, L. W. Molenkamp, Phys. Rev. B \textbf{62}, R7767 (2001).

\bibitem{Kusrayev2} Yu. G. Kusrayev, K. V. Kavokin, G. V. Astakhov, W. Ossau, L. W. Molenkamp, Phys. Rev. B \textbf{77}, 085205 (2008).

\end{references}
\end{document}